\documentclass[aps,prb,floatfix,twocolumn]{revtex4}
\usepackage{amsfonts}
\usepackage{amsmath}
\usepackage{amssymb}
\usepackage{graphicx}%
\begin{document}
\title{Non-collinear single-electron spin-valve transistors}
\author{Wouter Wetzels}
\author{Gerrit E.W. Bauer}
\affiliation{Kavli Institute of Nanoscience, Delft University of Technology, Lorentzweg 1,
2628 CJ Delft, The Netherlands}
\author{Milena Grifoni}
\affiliation{Institut f\"{u}r Theoretische Physik, Universit\"{a}t Regensburg, 93035
Regensburg, Germany}

\pacs{85.75.-d, 72.25.Mk, 73.23.Hk}

\begin{abstract}
We study interaction effects on transport through a small metallic cluster
connected to two ferromagnetic leads (a single-electron spin-valve transistor)
in the \textquotedblleft orthodox model\textquotedblright for the Coulomb
blockade. The non-local exchange between the spin accumulation on the island
and the ferromagnetic leads is shown to affect the transport properties such
as the electric current and spin-transfer torque as a function of the magnetic
configuration, gate voltage, and applied magnetic field.

\end{abstract}
\maketitle

\begin{figure}
\centering
\includegraphics[width=3.375in]{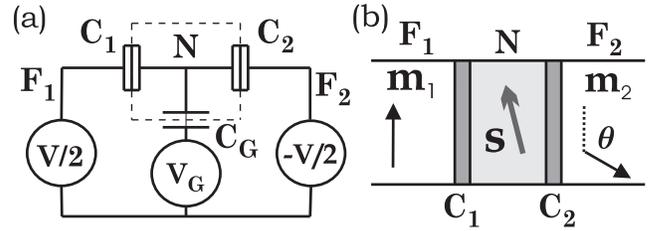}\caption{The spin-valve
single-electron transistor. (b) The tunneling rates between the leads and the cluster
depend on the spin accumulation $\vec{s}$.}%
\label{schematic}%
\end{figure}

Magnetoelectronics is a contender to fulfill the technological need for faster
and smaller memory and sensing devices. The drive into the nanometer regime
brings about an increasing importance
of electron-electron interaction effects. Small metallic clusters (islands)
that are electrically contacted to metallic leads by tunnel junctions and
capacitively coupled to a gate electrode can behave as \textquotedblleft single-electron transistors (SETs)\textquotedblright. In the Coulomb-blockade
regime, the charging energy needed to change the electron number on the island
by one exceeds the thermal energy and transport can be
controlled on the level of the elementary charge.\cite{Grabert Review SET} In
a spin-valve SET (SV-SET), the contacts to the cluster consist of
ferromagnetic metals (F). We focus here on F%
$\vert$%
N%
$\vert$%
F structures with normal metal (N) islands (see Fig.\ \ref{schematic}(a)),\cite{SV} since these \textquotedblleft spin
valves\textquotedblright display giant magnetoresistance and spin-current
induced magnetization reversal.\cite{Slonczewski} Other combinations such as F%
$\vert$%
F%
$\vert$%
F\cite{oso}, N%
$\vert$%
N%
$\vert$%
F\cite{Ralph} or F%
$\vert$%
F%
$\vert$%
N\cite{Yakushiji,Waintal} are of interest as well.

Several theoretical studies have been devoted to the binary magnetoresistance
(MR) of SV-SETs, \textit{i.e.} the difference in the electric resistance
between parallel and antiparallel configurations of the magnetization
directions.\cite{bf,BW} Interaction effects in magnetic devices have been
studied as well for spin valves with a Luttinger liquid island\cite{BE} and
for single-level quantum dots\cite{km,NC, braig} with non-collinear magnetic configurations.

A necessary condition for a significant MR in F%
$\vert$%
N%
$\vert$%
F structures is a spin accumulation on the normal metal island, \textit{viz}.
a sufficiently long spin-flip relaxation time $\tau_{\mathrm{sf}}$. Seneor
\textit{et al}.\cite{Seneor} have measured the MR of SV-SETs with
gold islands with a $\tau_{\mathrm{sf}}\sim800$ ps. A $\tau_{\mathrm{sf}}$ in
the microsecond regime has been reported for Co
nanoclusters.\cite{Yakushiji} The long spin-flip times
in small clusters is not yet fully understood; it might simply be due to the probability of finding zero impurities in a given small cluster.

In this Rapid Communication we elucidate the new aspects of electron transport
related to non-collinear magnetization directions in metallic SV-SETs. It
turns out that an effective exchange effect between the spin accumulation and
the magnetizations has to be taken into account.

We take the junction resistances sufficiently larger than the quantum
resistance $R_{Q}=h/e^{2}$ so that the Coulomb blockade can be treated by
lowest-order perturbation theory. We
furthermore disregard the size quantization of states in the clusters, thus
adopting the well-established \textquotedblleft orthodox\textquotedblright%
\ model.\cite{Grabert Review SET} In our model system (\textit{cf}. Fig.\ \ref{schematic}(b)), 
the ferromagnetic leads are treated as reservoirs with
single-domain magnetization directions $\vec{m}_{1}$ and $\vec{m}_{2}.$
Disregarding magnetic anisotropies, the relevant parameter is the angle
$\theta$ between the magnetizations. The capacitances of the junctions are
$C_{1}$ and $C_{2}.$ The cluster is capacitively coupled to the gate, with
capacitance $C_{G}\ll$ $C_{1},C_{2}$.

We assume a separation of time scales between the energy relaxation that
rapidly thermalizes injected charges and the slow spin relaxation. In this
regime, the quasi-equilibrium excess spin $\vec{s}$ on the normal metal island
is well-defined. In second quantization $\vec{s}=(\hbar/2)\sum_{kss^{\prime}%
}\left\langle c_{ks}^{\dagger}\vec{\sigma}_{ss^{\prime}}c_{ks^{\prime}%
}\right\rangle ,$ where $\vec{\sigma}$ is the vector of Pauli spin matrices
and $k$, $s$ denote the orbital and spin indices of the island states, respectively. This
corresponds to a chemical potential difference (spin accumulation) $\Delta
\mu=2\delta\left\vert \vec{s}\right\vert /\hbar,$ where $\delta$ is the
average single-particle energy separation (in terms of the static
susceptibility $\chi_{s}:$ $\Delta\mu=2\mu_{B}^{2}\left\vert \vec
{s}\right\vert /\left(  \chi_{s}\hbar\right)  $). Spin-flip relaxation is
parametrized by the spin-flip time $\tau_{sf}$ or spin-flip conductance
$G_{sf}\equiv e^{2}/\left(  2\tau_{sf}\delta\right)  $. For metals, $\delta$
is much smaller than the thermal energy except for very small particles
(diameter $\lesssim$5 nm). We restrict our attention here to a regime in which
the bias energy is small compared to the thermal and the charging energies,
but large compared to $\delta,$ so that the transport properties do not
depend on the energy-relaxation rate.\cite{BW} The
island state is then characterized by the excess number of electrons $n$ and net spin
angular momentum $\vec{s}$. A state distribution on the island governed
by free-energy minimization under these spin and charge constraints can be used.

The master equation for electron transport in the orthodox model is determined
by a tunneling Hamiltonian that can be treated by perturbation theory. We
derive the appropriate Hamiltonian by
collecting the leading terms in the transition probabilities. A crucial parameter is the mixing conductance for a N%
$\vert$%
F junction \cite{bnb2}
\begin{equation}
G^{\uparrow\downarrow}\equiv\frac{e^{2}}{h}%
{\displaystyle\sum\limits_{nm}}
\left(  \delta_{mn}-r_{\uparrow}^{nm}\left(  r_{\downarrow}^{nm}\right)
^{\ast}\right)  ,
\end{equation}
where $n$ and $m$ denote the transport channels in the normal metal, and
$r_{\uparrow}^{nm}$ and $r_{\downarrow}^{nm}$ are the corresponding
spin-dependent reflection coefficients. The real part of the mixing
conductance determines the spin-transfer torque that strives to align the
magnetization to the spin accumulation.\cite{Slonczewski} In the limit of a
tunneling contact, $\operatorname{Re}G^{\uparrow\downarrow}=\left(
G_{\uparrow}+G_{\downarrow}\right)  /2,$ where $G_{s}=\left(  e^{2}/h\right)
\sum_{nm}\left(  \delta_{mn}-\left\vert r_{s}^{nm}\right\vert ^{2}\right)  $
is the conventional tunneling conductance for spin $s$. The torque is
transferred by the electrons that tunnel through the contact and is included
in the standard spin-dependent tunneling Hamiltonian. The imaginary part of
the mixing conductance can be interpreted as an effective field parallel to
the magnetization direction of the ferromagnet. For a normal metal separated from a Stoner ferromagnet by a specular,
rectangular barrier we can directly solve the Schr\"{o}dinger equation. With
Fermi momenta of the ferromagnet of $k_{F\uparrow}=1.09\mathring{A}^{-1}$,
$k_{F\downarrow}=0.42\mathring{A}^{-1}$ (characteristic for Fe), a normal
metal Fermi energy of $\epsilon_{F}$ \thinspace$=2.6$ $eV$, a barrier height
of $3$ $eV,$ and free electron masses (cf.\ Ref. \onlinecite{Slonczewski}) we find $\operatorname{Im}%
G^{\uparrow\downarrow}/\left(
G_{\uparrow}+G_{\downarrow}\right)=-0.26$. This illustrates that, in contrast to
metallic interfaces,\cite{xia} the imaginary part of the mixing conductance
can be significant for tunnel junctions. This result is not sensitive to the
width of the barrier but may depend strongly on material combination and
interface morphology. For tunneling barriers made from magnetic
 insulators,\cite{EuO} $\operatorname{Re}G^{\uparrow\downarrow}$ and 
$\operatorname{Im}G^{\uparrow\downarrow}$ may become large compared to $(G_{\uparrow}+G_{\downarrow})$
,which results in different physics.\cite{Dani}

The Hamiltonian for the SV-SET reads:
\begin{equation}
H=H_{\mathrm{N}}+\sum\limits_{\alpha=1,2}\left(  H_{F\alpha}+H_{T\alpha
}+H_{\text{ex}\alpha}\right)  ,
\end{equation}
where $H_{N}$ describes the electrons on the normal metal and includes the
electrostatic interaction energy\cite{op}
\begin{equation}
H_{\mathrm{N}}=\sum_{ks}\varepsilon_{k}c_{ks}^{\dagger}c_{ks}+\frac
{e^{2}\left(  n-C_{G}V_{G}/e\right)  ^{2}}{2\left(  C_{1}+C_{2}\right)  }.
\end{equation}
For the two ferromagnetic leads ($\alpha=1,2$) $H_{F\alpha}=\sum
_{ks}\varepsilon_{\alpha ks}a_{\alpha ks}^{\dagger}a_{\alpha ks}$ and $n$
denotes the excess electron number on the island. The tunneling Hamiltonian
for each contact reads $H_{T\alpha}=\sum_{kqs}T_{\alpha kqs}a_{\alpha
ks}^{\dagger}c_{\alpha qs}+h.c.$ Finally, the imaginary part of the mixing
conductance gives rise to an effective exchange effect:
\begin{equation}
H_{\text{ex}\alpha}=\sum_{kss^{\prime}}\Delta_{\text{ex}\alpha}\vec{m}%
_{\alpha}\cdot c_{ks}^{\dagger}\vec{\sigma}_{ss^{\prime}}c_{ks^{\prime}%
},\label{ex}%
\end{equation}
where $\Delta_{\text{ex}\alpha}=-\hbar\delta\operatorname{Im}G_{\alpha
}^{\uparrow\downarrow}/\left(  2e^{2}\right)  .$ We note that the effect of
Eq.\ (\ref{ex}) on the spin accumulation is identical to that of an external
magnetic field applied in the direction $\vec{m}_{\alpha}$. Such an effective
exchange Hamiltonian has been introduced before for a Luttinger liquid
attached to ferromagnetic leads.\cite{BE} Physically, electrons in N feel the
ferromagnet through their tunneling tails that cause a spin-dependence of the
reflection coefficients. The small spin splitting due to
$\Delta_{\text{ex}}$ in the ground state does not influence the transport in the 
leading order of perturbation theory.

Recently, the angular dependence of transport through spin valves has been
studied for single-level quantum dot islands.\cite{km, braig} 
In these systems an effective field was found to act on unpaired quantum dot electron spins that
is caused by virtual particle exchange with the leads in the Coulomb blockade.
These correlations cause effects similar to those discussed here, but their
physical origin is completely different. Eq.\ (\ref{ex}) is caused by electron
exchange on a very fast time scale corresponding to the reciprocal Fermi
energy and reflects the electronic structures of the junctions, independent of
the applied voltages and charging energies. In contrast, the
correlation-mediated exchange is induced on time scales of the reciprocal
charging energy, changes sign with gate voltage and does not vanish for normal
metal contacts.\cite{braig} We find that also for classical islands the
correlation exchange field can be of the same order as $\Delta_{\text{ex}}$,
but since both effects can at least in principle be distinguished
experimentally by gate voltage and temperature dependence, a more detailed
discussion is deferred to a future publication.

We introduce spin-dependent conductances for both junctions, $G_{\alpha s}\equiv\pi
e^{2}\rho_{N}\rho_{F\alpha s}\overline{T}_{\alpha s}/\hbar$ ($\alpha=1,2,$
$s=\uparrow,\downarrow$). $\rho_{N}$ is the density of states at the Fermi
level in the normal metal, and $\rho_{F\alpha s}$ the spin-dependent density
of states in ferromagnet $\alpha$. $\overline{T}_{s}$ is proportional to the
average tunneling probability over all channels for spin $s$, $\overline
{T}_{s}\equiv\left\langle \left\vert T_{mns}\right\vert ^{2}\right\rangle
_{mn}.$ The conductances are assumed to be constant within the\ energy
interval of the charging energy, which is a safe assumption for
metals. We introduce the total conductances $G_{\alpha}\equiv G_{\alpha
\uparrow}+G_{\alpha\downarrow}$, the polarizations $P_{\alpha}\equiv\left(
G_{\alpha\uparrow}-G_{\alpha\downarrow}\right)  /G_{\alpha}$ and $F\left(
E\right)  \equiv E/\left(  1-e^{-\beta E}\right).$ The tunneling rate for adding
an electron through contact $\alpha\left(  =1,2=+,-\right)  $ in the considered regime where 
$eV \ll k_{B}T$ then reads:
\begin{gather}
\Gamma_{\alpha}^{n\rightarrow n+1}\left(  V,q,\vec{s}\right)  =\frac{G_{\alpha}}%
{e^{2}}F\left(  -E_{\alpha}\left(  V,q\right)  \right)  \nonumber\\
-\frac{G_{\alpha}}{e^{2}}F^{\prime}\left(  -E_{\alpha}\left(  V,q\right)
\right)  \frac{P_{\alpha}\Delta\mu}{2}\left(  \vec{m}_{\alpha}\cdot\widehat
{s}\right)  ,
\end{gather}
where $E_{\alpha}\left(  V,q\right)  =\alpha\kappa_{\alpha}eV-e\left(
q-e/2\right)  /\left(  C_{1}+C_{2}\right)  $ is the electrostatic energy
difference associated with the tunneling of one electron into the cluster from
lead $\alpha$ with $q=-ne+C_{G}V_{G}$ the charge on the island and
$\kappa_{\alpha}\equiv\left(  1/C_{1}+1/C_{2}\right)  ^{-1}/C_{\alpha}.$ The
other rates can be found analogously. We can also
find the net spin current into the cluster due to tunneling events in which an
electron is added to the island through contact $\alpha$:
\begin{align}
\left(  \frac{d\vec{s}}{dt}\right)  _{\alpha,\text{ tun}}^{n\rightarrow n+1}
&  =\frac{\hbar}{2}\frac{G_{\alpha}}{e^{2}}F\left(
-E_{\alpha}\left(  V,q\right)  \right)P_{\alpha}\vec{m}_{\alpha}  \nonumber\\
&  -\frac{\hbar}{2}\frac{G_{\alpha}}{e^{2}}F^{\prime}\left(  -E_{\alpha
}\left(  V,q\right)  \right)  \frac{\Delta\mu}{2}\widehat{s}.
\end{align}
The dynamics of the spin accumulation is also affected by spin-flip
scattering, the exchange effect discussed above and an external magnetic field
$\overrightarrow{B}$:%
\begin{align}
\left(  {\frac{d\vec{s}}{dt}}\right)  _{\text{sf}} &  =-\vec{s}/\tau_{sf},\\
\left(  {\frac{d\vec{s}}{dt}}\right)  _{\text{ex}} &  =\sum_{\alpha}%
\frac{\operatorname{Im}G_{\alpha}^{\uparrow\downarrow}\delta}{e^{2}}\left(
\vec{m}_{\alpha}\times\vec{s}\right),  \\
\left(  {\frac{d\vec{s}}{dt}}\right)  _{\text{magn}} &  =\frac{g\mu_{B}%
}{\hbar}\left(  \overrightarrow{B}\times\vec{s}\right).
\end{align}
The master equation for the SV-SET is $dp_{n}/dt=-p_{n}\left(  \Gamma
^{n\rightarrow n+1}+\Gamma^{n\rightarrow n-1}\right)  +p_{n+1}\Gamma
^{n+1\rightarrow n}+p_{n-1}\Gamma^{n-1\rightarrow n}$ combined with $d\vec
{s}/dt\ $described above. Here, $p_{n}$ is the probability distribution for
the number of electrons on the island and $\Gamma^{n\rightarrow n+1}$ denotes
the rate for adding another electron when the cluster has $n$ excess
electrons. We now focus on a quasi-stationary state $dp_{n}/dt=0$. When $eV\ll k_{B}T\ll e^{2}/2C$
the number of electrons on the island fluctuates between two values denoted
by $n=``0"$ and $``1"$. We can use
detailed balance symmetry to find the stationary state, whence $p_{0}%
\Gamma^{0\rightarrow1}=p_{1}\Gamma^{1\rightarrow0}$. We find the conductance $G$:
\begin{equation}
G\left(  V_{G},\vec{s}\right) =G_{\text{KS}}\left(V_{G}\right) \left[ 1+\frac{\Delta\mu}{2eV}\left(  P_{1}\vec{m}_{1}%
-P_{2}\vec{m}_{2}\right)  \cdot\vec{s}/\left\vert \vec{s}\right\vert \right]
,
\end{equation}
where $G_{\text{KS}}\left(V_{G}\right)\equiv\beta\Delta_{\min}G_{1}G_{2}/\left[  2\left(
G_{1}+G_{2}\right)  \sinh\beta\Delta_{\min}\right] $ with $\Delta_{\min}$
$\equiv e\left(  C_{G}V_{G}-e/2\right)  /\left(  C_{1}+C_{2}\right)  $) describes one
Coulomb blockade oscillation.\cite{Kulik and Shekhter}

The spin accumulation is obtained from the condition $d\vec{s}/dt=0.$ For a
symmetric setup, in which the conductance parameters that characterize the two
contacts are equal, we find:
\begin{equation}
\frac{G\left(V_{G}\right)  }{G_{\text{KS}}\left(V_{G}\right)}=1-\frac{P^{2}}{1+\frac{G_{sf}%
}{G_{\text{KS}}\left(V_{G}\right)}}\frac{\tan^{2}\frac{\theta}{2}}{\tan^{2}\frac{\theta}%
{2}+1+\left[  \frac{\operatorname{Im}G^{\uparrow\downarrow}}{G_{\text{KS}%
}\left(V_{G}\right)+G_{sf}}\right]  ^{2}}
\end{equation}
When the imaginary part of the mixing conductance vanishes, the angular
dependence of the conductance is proportional to a cosine. The result of
Brataas \textit{et al.} for a F%
$\vert$%
N%
$\vert$%
F\ spin valve without interaction\cite{bnb2} is recovered by substituting
$\beta\Delta_{\min}/\left(  2\sinh\beta\Delta_{\min}\right)$ by 1 
(cf.\ Ref.\ \onlinecite{BW}).

\begin{figure}
\centering
\includegraphics[width=3.375in]{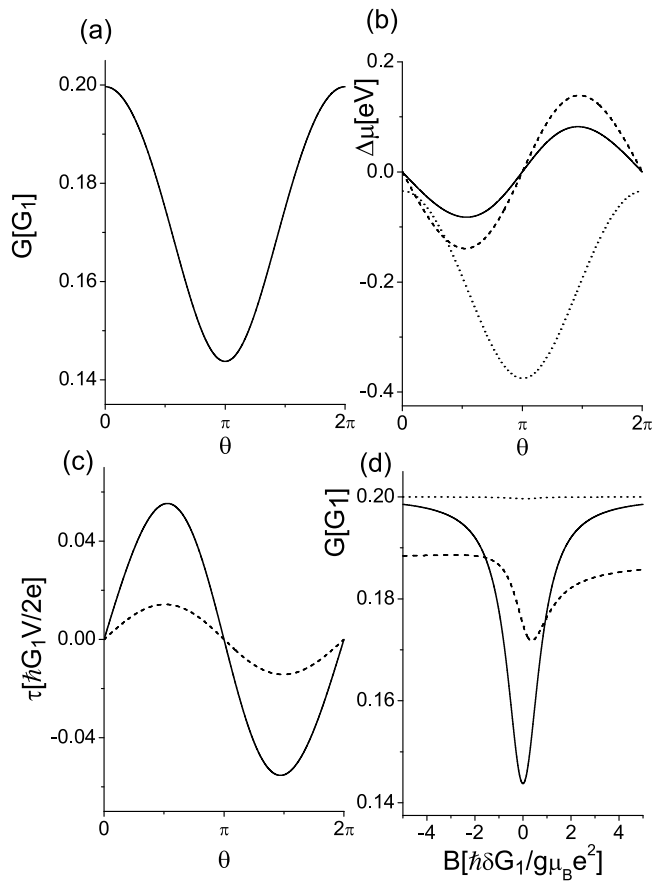}\caption{Results for a SV-SET with 
$G_{2}=2 G_{1}/3$, $P_{1}=0.8$, $P_{2}=0.7,$ $G_{sf}= G_{1}/5,$
$\operatorname{Im}G_{1}^{\uparrow\downarrow}=G_{1}/4,$ $\operatorname{Im}%
G_{2}^{\uparrow\downarrow}=G_{2}/4$ and $\Delta_{\min}=0$. (a) Conductance as a function of the
angle between the magnetizations of the leads. (b) Spin accumulation components in the
direction of $\vec{m}_{1}$ (dotted), $\vec{z}$ $\times\vec{m}_{1}$ (dashed),
and $\vec{z}$ (solid). (c) Spin transfer torque on ferromagnet 1 in the
direction $\vec{z}\times\vec{m}_{1}$ (dashed) and $\vec{z}$ (solid). (d) Conductance 
\textit{vs}.\ magnetic field applied in the direction $\vec{z}$ $\times\vec{m}_{1}$ for 
$\theta=0$ (dotted), $\theta=\pi/2$ (dashed) and $\theta=\pi$ (solid).} 
\label{Angular dependence}%
\end{figure}

In Fig.\ \ref{Angular dependence}(a) an example of the dependence of the
conductance on the angle $\theta$ is shown. When the magnetizations are
non-collinear, the exchange effect reduces the spin accumulation and increases
the conductance. In Fig.\ \ref{Angular dependence}(b), the spin accumulation as
a function of angle illustrates that for non-collinear magnetizations, the
non-local exchange pulls the spin accumulation vector out of the plane of the
magnetizations. 

The spin current between the ferromagnets and the normal metal
gives rise to a spin-transfer torque 
$\tau_{\alpha}=-\vec{m}_{\alpha} \times \left( {\bf{I}}_{s} \times \vec{m}_{\alpha}\right)$ on ferromagnet $\alpha$
 (see Fig.\ \ref{Angular dependence}(c)). ${\bf{I}}_{s}$
is the net spin flow out of the ferromagnet,  
and is strongly modulated by the gate voltage. When the Coulomb interaction
suppresses the current, the exchange effect becomes relatively more important. 
Fig.\ \ref{Angular dependence}(d)) shows that, for non-collinear configurations, the exchange effect 
causes an asymmetry in the Hanle effect with respect to the sign of an applied external magnetic field.

\begin{figure}
\centering
\includegraphics[width=3.375in]{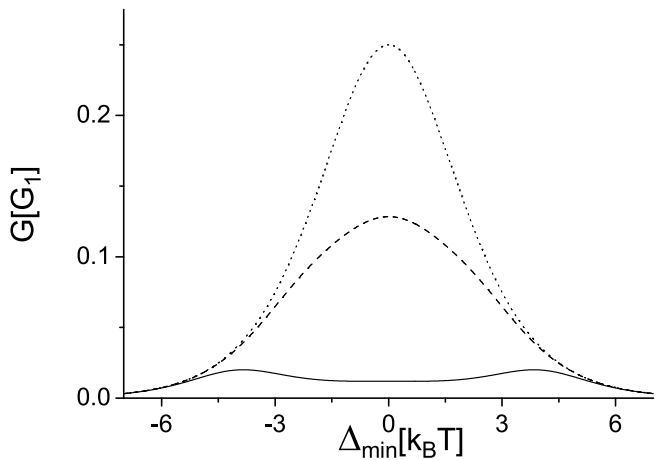}\caption{Conductance (G) as a function 
of gate voltage for $\theta=0$ (dotted), $\theta=0.6\pi$ (dashed), and $\theta=0.9\pi$
(solid). The ferromagnets are half-metallic and the conductance of both
contacts is $G_{1}$. $\operatorname{Im}G_{1}^{\uparrow\downarrow}%
=\operatorname{Im}G_{2}^{\uparrow\downarrow}=G_{1}/4,$ $G_{sf}=0.$}%
\label{Coulomb oscillation}%
\end{figure}

The curves in Fig.\ \ref{Coulomb oscillation} show the gate-voltage modulated
conductance for a symmetric spin valve with
half-metallic ferromagnetic leads ($P=1$). Spin-flip is disregarded. When
the magnetizations are parallel ($\theta=0$), no spin accumulates on the
island and the Coulomb-blockade oscillation equals that of all-normal metal
systems. When the angle $\theta$ is increased, the conductance is suppressed by a
counteracting spin accumulation. The exchange effect acts to reduce this spin accumulation 
for non-collinear configurations and can cause a local conductance minimum at $\Delta_{\min}=0$ (e.g. for 
$\theta=0.9$ $\pi)$. Deep in the Coulomb blockade, a significant spin accumulation 
is prevented from building up and all curves converge.  

Interestingly, the angular magnetoresistance for Luttinger liquids with
ferromagnetic contacts\cite{BE} looks very similar. In order to find
experimental evidence for spin-charge separation it is therefore necessary to
avoid spurious effects caused by the Coulomb blockade.

In conclusion, we studied the transport characteristics for non-collinear spin
valves in the Coulomb blockade regime. A non-local exchange interaction
between the spin accumulation and the ferromagnets  affects the conductance
and spin-transfer torque as a function of the gate voltage. This might provide
new possibilities to control charge and spin transport in nanoscale
magnetoelectronic devices.

We acknowledge valuable discussions with Yuli Nazarov, Jan Martinek,
J\"{u}rgen K\"{o}nig and Yaroslav Tserkovnyak. The research was supported by
the NWO, FOM, and EU Commission FP6 NMP-3 project 05587-1 \textquotedblleft
SFINX\textquotedblright.


\begin{thebibliography}{99}
\bibitem {op}Stricly speaking $n=%
{\textstyle\sum\nolimits_{ks}}
c_{ks}^{\dagger}c_{ks}$ is here an operator.
\end{thebibliography}
\end{document}